\documentclass[prd,twocolumn,,floatfix,amsmath,amssymb]{revtex4-1}
\usepackage{amsmath,amsfonts,amsthm,amssymb,slashed,graphicx}
\usepackage{xcolor}
\usepackage{graphicx}
\usepackage{epstopdf}
\usepackage{array,booktabs}
\usepackage{hyperref}
\usepackage{bbm} % for \mathbbm{1}
\usepackage{mathtools} % for \mathclap
\usepackage{tabulary}
\usepackage{multirow}
\usepackage{array,arydshln}
\hypersetup{linktoc = all}  %hyperref settings
\hypersetup{pdfborderstyle={/S/U/W 0.5}}
\usepackage{tikz}
\usepackage{tkz-tab}
\usepackage{caption}
\usepackage{subcaption}
\usetikzlibrary{shapes.geometric}
\usetikzlibrary{intersections}
\usetikzlibrary{calc}
\usetikzlibrary{decorations.pathmorphing}
\newcounter{mycnt}
\newcommand{\be}{\begin{equation}}
\newcommand{\ee}{\end{equation}}
\newcommand{\ba}{\begin{eqnarray}}
\newcommand{\ea}{\end{eqnarray}}

\newcommand{\tcr}{\textcolor{red}}
\newcommand{\tcb}{\textcolor{blue}}

\begin{document}

\title{Higher Order Corrections to Holographic Black Hole Chemistry}

\author{Musema Sinamuli}
\email{cmusema@perimeterinstitute.ca}
\affiliation{Department of Physics and Astronomy, University of Waterloo, Waterloo, Ontario N2L 3G1, Canada}
\affiliation{Perimeter Institute for Theoretical Physics, Waterloo, Ontario, N2L 2Y5, Canada}

\author{Robert B. Mann}
\email{rbmann@uwaterloo.ca}
\affiliation{Department of Physics and Astronomy, University of Waterloo, Waterloo, Ontario N2L 3G1, Canada}

\begin{abstract}
We investigate the holographic Smarr relation beyond the large $N$ limit. By making use  of the holographic dictionary, we find that the bulk correlates of sub-leading $1/N$ corrections to this relation are related to the couplings in Lovelock gravity theories. We likewise obtain
a holographic equation of state, and check its validity for a variety of interesting and non-trivial black holes, including  rotating planar black holes in Gauss-Bonnet-Born-Infeld gravity, and non-extremal rotating black holes in minimal $5d$ gauged supergravity.
 We provide an explanation of the $N$-dependence of  the holographic Smarr relation in terms of contributions due to planar and non-planar diagrams in the dual theory.
 \end{abstract}

\maketitle
\section{Introduction}

For nearly two decades  the AdS/CFT correspondence \cite{thelarge} has been the subject of intense research, motivated by the fact that it posits a connection between an anti de Sitter (AdS) black hole and a conformal field theory (CFT) defined on its boundary. In the
context of black hole thermodynamics, an
 understanding of the physics of the AdS black hole can be reinterpreted  in terms of a thermal system on the boundary field theory and vice-versa. 
 
The general assumption underlying nearly all investigations of the AdS/CFT correspondence is that the cosmological is a fixed parameter. 
However increasing interest has been focussed on regarding  the cosmological constant as a thermodynamic variable via the relation 
\begin{equation}\label{plam}
P = -\frac{\Lambda}{8\pi G_{d}} = \frac{(d-1)(d-2)}{16\pi l^2 G_{d}}
\end{equation}
where $P$ is the pressure of the black hole system \cite{Kastor:2009wy}. 
 With this comes the associated concept of volume $V$ for a black hole, which is the
thermodynamic conjugate of pressure.  The extension of thermodynamic phase space to include these two variables has led to the realization that black holes can exhibit enormously rich and diverse  phase behaviour, including Van der Waals phase transitions 
for charged black holes \cite{Kubiznak:2012wp,Gunasekaran:2012dq},   triple points analogous to water 
\cite{Altamirano:2013uqa}, re-entrant phase transitions analogous to those seen in gels and polymers \cite{Altamirano:2013ane}, and even superfluid phase transitions analogous to those in superfluid helium \cite{Hennigar:2016xwd}.
This burgeoning subfield is now referred to as  black hole chemistry \cite{Kubiznak:2016qmn}

It is therefore of interest to ask what black hole chemistry implies for the variables in the boundary field theory. What do the first law of thermodynamics, the Smarr relation, and so on look like on the CFT side? The quantity $l$ in \eqref{plam} is the AdS radius and is
related to the number of colours $N$ in the dual gauge theory via a holographic relation of the form \cite{Karch:2015rpa}
\begin{equation}
\label{correspond1}
\frac{l^{d-2}}{G_d}\sim N^2
\end{equation}
where the $d$-dimensional gravitational constant $G_d$ has a length dimension of $d-2$. This kind of relation was first introduced in the AdS/CFT correspondence from string theory \cite{thelarge}, in which an $\mbox{AdS}_5\times S^5$ spacetime appears to be the near horizon geometry of $N$ coincident $D_3$ branes in type IIB supergravity. The correspondence between an $\mbox{AdS}_5\times S^5$ spacetime and a ${\cal{N}}=4$ SU(N) Yang-Mills theory on its boundary was expressed as follows
\begin{equation}
l^4=\frac{\sqrt{2}\ell_{Pl}^4}{\pi^2}N\,,
\label{eq:NrhL}
\end{equation}
where $\ell_{Pl}$ is the 10-dimensional Planck length. From the two preceding relations we can remark that the variation of the AdS radius $l$ amounts to the variation of the color number $N$ in the boundary Yang-Mills theory.
\vskip 5pt An interesting subject to think about is on the nature of the connection between the bulk and the CFT on its boundary as well as its implications when we are in presence of another theory of gravity.

The suggestion that varying the pressure, or $\Lambda$, is equivalent to varying the number of colors, $N$, in the boundary Yang--Mills theory has been proposed by
a few authors \cite{Johnson:2014yja,Dolan:2014cja,Kastor:2014dra}, with $V$ being interpreted in the boundary field theory as an associated chemical potential $\mu$ for colour.  This has the consequence that variation of $\Lambda$ in the bulk moves one around the space of field theories in the boundary. Alternatively, one could keep $N$ fixed, so that field theory remains the same, in which case varying $\Lambda$ in the bulk corresponds to varying the curvature radius governing the space on which the field theory is defined \cite{Karch:2015rpa}.
 
From this latter perspective, a generalized Smarr relation can be derived by considering the thermodynamics of the dual field theory  \cite{Karch:2015rpa}.  Noting that  the free energy of the field theory scales simply as $N^{2}$, we have
\begin{equation}\label{holosmarr}
\Omega(N,\mu,T,l)= N^{2}\Omega_{0}(\mu,T,l)  
\end{equation}
in the limit of  large $N$.  For a conformal field theory the equation of state reads
\be\label{UCFT}
E=(d-2)p{\cal V}
\ee
and together with \eqref{holosmarr} can be used to obtain the standard Smarr relation $(d-3)M = (d-2)TS-2PV$ for an uncharged AdS black hole.  In this sense equation \eqref{holosmarr} can be regarded as a 
`{\it holographic Smarr relation}'.  Since   $N^{2} \sim \frac{l^{d-2}}{G_d}$, varying $\Lambda$ is equivalent to varying the AdS length $l$,
and since $N$ is fixed,  $G_d$ must also be varied.

The purpose of this paper is to investigate the holographic Smarr relation \eqref{holosmarr} beyond the large $N$ limit, including sub-leading corrections to this relation.  We will see that relation (\ref{holosmarr}) can be generalized to a form that includes subleading $1/N$ corrections
whose bulk correlates are related to the couplings in Lovelock gravity theories.  Lovelock theories are higher curvature or derivative generalizations of Einstein's theory, and in the context of string theory are understood as quantum corrections to  Einstein gravity.  We
shall show that the Lovelock couplings are related  to a function of   $N$, with variations of the Lovelock couplings in the bulk dictating the behaviour of the corresponding CFT via the variation of these functions.

 Inquiries on how far  this generalization extends (or what are the limits thereof) for a given black hole are also be of considerable interest. The bulk Smarr relation and the corresponding CFT equation of state  are both expected to be satisfied at the lowest order (Einstein-Hilbert action). We shall look at what happens at higher order, considering especially the CFT equation of state
 to see whether or not it breaks down.

In section 2, we review some important notions and relations for Lovelock black holes, particularly the first law of thermodynamics and the Smarr relation. In section 3 we investigate how these important relations in the bulk theory are viewed in the CFT, particularly with regards to the derivation of the  equation of state in the boundary field theory. 
Section 4 is devoted to the holographic derivation of the Smarr relation where we mostly make use of the equation (\ref{extendedholo1}) of the holographic dictionary and by regarding the grand canonical function $\Omega$ as a homogeneous function of functions of $N$. 
In section 5 we check the validity of the equation of state, introduced earlier in section 3, for some particular cases of black holes, including  spherically symmetric AdS lovelock black holes,   rotating planar black holes in Gauss-Bonnet-Born-Infeld gravity, and   non-extremal rotating black holes in minimal $5d$ gauged supergravity.  An explanation of the dependence of the function of  $N$ is in
section 6, and  in the last section we make some concluding remarks.

\section{A Review of Lovelock black holes}
 
In this current section we review the derivation of thermodynamic quantities associated with   Lovelock black holes and some relations implied by these quantities.    

 Lovelock gravity is a generalization of Einstein's theory whose action and field equations are nonlinear in the curvature whilst always maintaining second-order differential equations for the metric.
  Its Lagrangian has the form \cite{theeinstein}
\begin{equation}
\label{lagrangian1}
L=\frac{1}{16\pi G_d}\sum^{\frac{d-1}{2}}_{k=0}\hat{\alpha}_{(k)}L^{(k)}
\end{equation}
with $d$ the spacetime dimension, $\hat{\alpha}_{(k)}$   the Lovelock coupling constants for the $k$-th power of curvature, and $L^{(k)}$ the Euler density of dimension $2k$. These Euler densities are expressed as
\begin{equation}
\label{lagrangian2}
L^{(k)}=\frac{1}{2^k}\delta^{a_1b_1...a_kb_k}_{c_1d_1...c_kd_k}R_{a_1b_1}^{c_1d_1}...R_{a_kb_k}^{c_kd_k}
\end{equation}
where the $\delta^{a_1b_1...a_kb_k}_{c_1d_1...c_kd_k}$ are the totally antisymmetric in both set of indices of the Kronecker delta functions and $R_{a_kb_k}^{c_kd_k}$ the Riemann tensors.

From the Lagrangian (\ref{lagrangian1}, \ref{lagrangian2}), the variational principle yields the vacuum equations of motion for Lovelock gravity,
which are
\begin{equation}
{\mathcal{G}}^{a}_{b}=\sum^{\frac{d-1}{2}}_{k=0}\hat{\alpha}_{(k)}{G^{(k)}}^{a}_{b}=0
\end{equation}
with ${G^{(k)}}^{a}_{b}$ the Einstein-like tensors, which read as
\begin{equation}
{G^{(k)}}^{a}_{b}=-\frac{1}{2^{k+1}}\delta^{aa_1b_1...a_kb_k}_{bc_1d_1...c_kd_k}R_{a_1b_1}^{c_1d_1}...R_{a_kb_k}^{c_kd_k}
\end{equation}
and each of them satisfy independently the conservation law $\nabla_a {G^{(k)}}^{a}_{b}=0$.

If we minimally couple to a Maxwell field $F_{ab}$ the action is
\begin{equation}
S=\frac{1}{16\pi G_d}\int d^dx\sqrt{-g}\big[\sum^{\frac{d-1}{2}}_{k=0}\hat{\alpha}_{(k)}L^{(k)}-4\pi G_dF_{ab}F^{ab}\big].
\end{equation}
and yields 
\begin{equation}
\label{einsteinmaxwell}
\sum^{\frac{d-1}{2}}_{k=0}\hat{\alpha}_{(k)}{G^{(k)}}^{a}_{b}=8\pi G_d\big[F_{ac}F_b^c-\frac{1}{4}g_{ab}F_{cd}F^{cd}\big]
\end{equation}
for the equations of motion. Without solving these equations, it can be shown for solutions of asymptotic constant curvature that
the first law of thermodynamics and the Smarr relation respectively are
\begin{eqnarray}\label{flaw}
\delta M&=&T\delta S 
+\mu\delta Q-\frac{1}{16\pi G_d}\sum^{\frac{d-1}{2}}_{k=0}\Psi^{(k)}\delta\hat{\alpha}_{(k)} \nonumber\\
(d-3)M&=&(d-2)TS+(d-3)\mu Q  \nonumber \\
&&\quad + \sum^{\frac{d-1}{2}}_{k=0}\frac{2(k-1)}{16\pi G_d}\Psi^{(k)}\hat{\alpha}_{(k)}
\label{smarr}
\end{eqnarray}
where the solution is characterized by a mass $M$, a charge $Q$,  Lovelock coupling constants $\hat{\alpha}_{(k)}$ each
having  thermodynamic conjugate $\Psi^{(k)}$,  and  (if it is a black hole) a temperature $T$ , and an entropy $S$.

 Restricting attention to   spherically symmetric metrics
\begin{eqnarray}
\label{sphericalsym}
ds^2&=&-f(r)dt^2+f(r)^{-1}dr^2+r^2d\Omega^{2}_{(\kappa)d-2}\nonumber\\
F&=&\frac{Q}{r^{d-2}}dt\wedge dr
\end{eqnarray}
where $d\Omega^{2}_{(\kappa)d-2}$ is the line element of a compact space of dimension $(d-2)$ with constant curvature $(d-2)(d-3)\kappa$
($\kappa=-1,0,1$), 
the equations of motion (\ref{einsteinmaxwell}) for   charged spherically symmetric black holes of mass $M$ read \cite{stringgenerated, symmetricsol1, symmetricsol2, anoteon, ontheuniversality, alovelockblack, pathologies}
\begin{eqnarray}
\sum^{\frac{d-1}{2}}_{k=0}\alpha_k\bigg(\frac{\kappa-f}{r^2}\bigg)^k&=&\frac{16\pi G_d M}{(d-2)\omega^{\kappa}_{d-2}r^{d-1}}-\frac{8\pi G_d Q^2}{(d-2)(d-3)r^{2(d-2)}}\nonumber\\
\label{f-def}
\end{eqnarray}
where the charge is given by
\begin{equation}
Q=\frac{1}{2\omega^{(\kappa)}_{d-2}}\int\ast F
\end{equation}
and where~~ $\omega^{(1)}_{d-2}=\frac{2\pi^{(d-1)/2}}{\Gamma((d-1)/2)}$~~ and 
\begin{eqnarray}
\alpha_0&=&\frac{\hat{\alpha}_{(0)}}{(d-1)(d-2)},~~~~~~~~\alpha_1=\hat{\alpha}_{(1)}\nonumber\\
\alpha_k&=&\hat{\alpha}_{(k)}\prod^{2k}_{n=3}(d-n)~~~~~~ \mbox{for}~~k\geq 2
\end{eqnarray}
is a simple and useful rescaling of the Lovelock couplings.
% From now on and for further purposes we will only consider spaces with $\kappa=1$. 

Computing the temperature $T=\frac{f^\prime(r_+)}{4\pi}$ via standard Wick rotation arguments, we need not explicitly know $f(r)$ in order to determine the mass $M$, temperature $T$, entropy $S$ and electric potential $\mu$ of the black holes. These thermodynamic quantities are \cite{anoteon,blackholeentropy}
\begin{eqnarray}
M&=&\frac{\omega^{(\kappa)}_{d-2}(d-2)}{16\pi G_d}\sum_{k=0}\alpha_k\kappa^k{r_+}^{d-1-2k}+\frac{\omega^{(\kappa)}_{d-2}Q^2}{2(d-3){r_+}^{d-3}}\nonumber\\
T&=&\frac{1}{4\pi r_+D(r_+)}\bigg[\sum_{k=0}\kappa\alpha_k(d-2k-1)(\frac{\kappa}{r_+^2})^{k-1}\nonumber\\
&&\qquad\qquad\qquad\qquad - \frac{8\pi G_dQ^2}{(d-2)r_+^{2(d-3)}}\bigg] 
\label{thermoquantities} \\
S&=&\frac{\omega^{(\kappa)}_{d-2}(d-2)}{4G_d}\sum_{k=0}\frac{k\kappa^{k-1}\alpha_kr_+^{d-2k}}{d-2k},~\mu=\frac{\omega^{(\kappa)}_{d-2}Q}{(d-3)r_+^{d-3}}
\nonumber
\end{eqnarray}
with $D(r_+)=\sum_{k=1}k\alpha_k(\kappa r_+^{-2})^{k-1}$ and $r_+$   the horizon radius.  From the extended first law \eqref{flaw} and Smarr relation
\eqref{smarr} we can obtain the thermodynamic conjugate quantities $\Psi^{(k)}$ \cite{multiple, isolatedcritical}
\begin{equation}
\label{conjugate}
\Psi^{(k)}=\frac{\omega^{(\kappa)}_{d-2}(d-2)}{16\pi G_d}r_+^{d-2k}\bigg[\frac{\kappa}{r_+}-\frac{4\pi kT}{d-2k}\bigg], ~~~~~k\geq 0
\end{equation}
in terms of the rescaled coupling constants.
The above quantities satisfy the Smarr relation (\ref{smarr}).  From these quantities it follows that the thermodynamic pressure and volume are given by
\begin{equation}\label{PandV}
P=-\frac{\Lambda}{8\pi G_d} = \frac{(d-1)(d-2)}{16\pi G_d}\alpha_0~~~~~~ V=\omega^{(\kappa)}_n ~\frac{r_+^{n+1}}{d-1}
\end{equation}
where $n=d-2$. 
 
  Before proceeding we pause to comment on the relationship between these quantities and the more standard notions of thermodynamics bulk  pressure
and volume in black hole thermodynamics, which are  \cite{smarrformulaand}
\begin{equation}\label{PandV}
P_b=-\frac{\Lambda}{8\pi}~~~\mbox{and}~~~ V_b=\frac{\partial M}{\partial P_b}\big|_{S,Q_b,\alpha\geq 1}.
\end{equation}
The first relation is the standard identification of $\alpha_0$ with the cosmological constant 
\cite{multiple}.
Hence
\begin{equation}
\label{pressure}
P_bV_b=\alpha_0\Psi^{(0)}.
\end{equation}
The CFT pressure and volume can be defined as $p$ and $v=\omega_n^{(\kappa)}R^n$. The pressure $p$ has a length dimension of $-(n+1)$ and $R$ is the radius of the sphere on which the CFT is defined.

\section{Equation of state}

In this section we   derive the equation of state by looking at how transformations of parameters on the field theory lead to transformations in the bulk or vice versa. A example of this is the proposed correspondence between varying the cosmological constant $\Lambda\sim(\alpha_0)$  in the bulk and  variations in the number of colors $N$ in the field theory  \cite{Kastor:2014dra, Johnson:2014yja, phasetrans1, phasetrans2, holoentang}.   

The   Smarr relation \eqref{smarr} has been posited \cite{Karch:2015rpa} to be derivable from the scaling
properties of the free energy of the dual field theory   in the limit of a large number of colors $N$. The free energy 
$\Omega(N,\mu,T,l)$
of the field theory 
dual to Einstein-AdS gravity
scales as
\begin{equation}\label{holosmarr}
\Omega(N,\mu,T,l)= N^{2}\Omega_{0}(\mu,T,l) \,.
\end{equation}
where $N^{2}$ is the central charge.  
%\tcr{\bf [Check this.]}\tcb{\bf [Correct!]}
Extending $\Omega(N,\mu,T,l)$ to   Lovelock gravity theory we posit
\begin{equation}\label{felov}
\Omega(N, \mu, T, \alpha_j,  R)=\sum_{k=0}g_k(N)\Omega^k(\mu, T, \alpha_j,  R)
\end{equation}
%where   $R$ ~\tcb{the radius of the sphere on which the field theory dwells}. 
where the $g_k(N)$ are assumed to be polynomial functions (as suggested in \cite{Karch:2015rpa}) 
of $N$.  We will see in the next section that this form is of great interest in the derivation of the holographic Smarr relation
for Lovelock gravity.

Noting that the thermal properties of AdS black holes  can be reinterpreted as those of a CFT  at the same finite temperature \cite{Witten}, the grand canonical free energy and its density are expressible in terms of the on-shell action the (Euclidean) bulk solution as \cite{Karch:2015rpa}
\begin{equation}\label{Omega-eq}
\Omega=M-TS-\mu Q~~~~~~~~~\tilde{\Omega}=\tilde{M}-T\tilde{S}-\mu\tilde{Q}
\end{equation}
where the quantities $\tilde{M}, \tilde{S}$ and $\tilde{Q}$ are the respective mass,   entropy and  charge per unit volume of the CFT. Note  that these thermodynamic quantities are defined on the boundary and have the following form: $Q\sim Q_bl,~\mu\sim\mu_b/l,~~\alpha_k^F=\alpha_k l^{2(1-k)}$ and $\Psi_F^{(k)}=\Psi^{(k)}l^{2(k-1)}$, while others are kept unchanged.

Let us consider conformal field theories, whose equations of state is obtained by taking into account the behaviour of the thermodynamic quantities under an infinitesimal scale transformation 
\begin{eqnarray}
\label{transformation}
dS&=&0,~~dQ=0,~~ d\alpha_k^F=0~~(k\geq 1),~~dM=Md\lambda\nonumber\\
dp&=&(n+1)pd\lambda ~~~~~dv=-nvd\lambda
\end{eqnarray}
$\lambda$ is the parameter associated to the scale transformation. 
From these relations and the extended first law of thermodynamics
\begin{equation}
\label{firstlaw}
 dM= T dS+\mu dQ+ \sum_{k}\Psi_F^{(k)}d\alpha_k^F
\end{equation}
we are led to the equation of state
\begin{equation}
\label{equationofstate1}
\tilde{M}=(n+1)p.
\end{equation}
%\tcr{\bf [Are you sure that it is $(n+1)$ and not $n$?  Compare to (5.26) of 1608.06147]}\tcb{\bf [ I am very sure. It comes from eq (\ref{transformation}). This is congruent with the definition of CFT pressure which goes like $l^{-(n+1)}$ .}
Also the extended first law is reduced to
\begin{equation}
dM=vdp 
\end{equation}
and (under the constraints of \eqref{transformation}) knowing that 
\begin{eqnarray}
v&=&\frac{\partial\Omega}{\partial p}  = (\tilde{\Omega} \partial_Rv+v\partial_R\tilde{\Omega})\frac{\partial R}{\partial p}
\end{eqnarray}
we get  
\begin{equation}\label{dRp}
\partial_Rp=\frac{n}{R}\tilde{\Omega}+\partial_R\tilde{\Omega}
\rightarrow
n\tilde{\Omega}+R\partial_R\tilde{\Omega} = -(n+1)p
\end{equation}
where $R\partial_Rp=-(n+1)p$ because of the length dimension of $p$.
%Since the pressure as defined in (\ref{pressure}) has a length dimension of $-(n+1)$, it is easy to show that it satisfies the equation  \tcb{( when we assume that $\lambda$ in (\ref{transformation}) scales like $\lambda\sim -\log R$)}. 
Inserting the equation of state (\ref{equationofstate1}) into \eqref{dRp} yields
\begin{equation}
\tilde{M} = -(n\tilde{\Omega}+R\partial_R\tilde{\Omega}) 
\end{equation}
or alternatively, using \eqref{Omega-eq},
\begin{equation}
\label{equationofstate2}
(n+1)\tilde{M}=n(T\tilde{S}+\mu\tilde{Q})-R\partial_R\tilde{\Omega}
\end{equation}
which is the holographic equation of state.

For rotating black holes (\ref{equationofstate2}) has to be slightly modified; we have to add one more condition to (\ref{transformation})
\begin{equation}
dJ_i= J_i d\lambda
\end{equation}
with $J_i$ the angular momentum associated to the i-th angular variable. 
The additional condition takes (\ref{equationofstate2}) to the new expression 
\begin{equation}
\label{newstate}
(n+1)\tilde{M}=n(T\tilde{S}+\mu\tilde{Q})+(n+1)\sum_i\omega_i \tilde{J}_i-Rd_R\tilde{\Omega}
\end{equation}
where $\tilde{\Omega}_i$ is the angular velocity associated with the i-th angular variable and~ $\tilde{\Omega}= \tilde{M}-T\tilde{S}-\mu\tilde{Q}-\sum_i\omega_i\tilde{J}_i$ .

\section{Holographic Smarr relation}

%In the previous section we gave few details of the form of the grand canonical free energy as a function of the color number $N$ of the field theory and other thermodynamic quantities associated to the bulk. In the next few paragraphs we will have the task to provide more explanation on the choice of such a form.   

 The grand canonical free energy \eqref{felov}  introduced in the previous section is a polynomial on the variable $N^2$.  In Einstein gravity   \cite{Karch:2015rpa} only the first term of $\Omega$ is taken into account. This can be justified by the fact that the dual field theories to the black holes are considered to be in the large $N$ limit.
% \tcr{\bf [Does the preceding sentence make sense?]}\tcb{\bf [Corrected.]}

For a Lovelock black hole nonzero additional terms appear   due to contributions from the higher curvature terms. Without knowing explicitly their form, the dimensionality 
\begin{equation}
\label{extendedholo2}
[\alpha_k]=2(k-1)~~~~~~~\mbox{or}~~~~~\alpha_k\sim l^{2(k-1)} 
\end{equation}
of the Lovelock couplings implies that 
\begin{equation}
\label{extendedholo1}
\beta_k(\alpha_k)^{\frac{d-2}{2(k-1)}}=g_k(N)
\end{equation}
 for which the $k=0$ term is  
\begin{equation}
\beta_0 l^{d-2}=N^2
\end{equation}
recovering the relationship \eqref{correspond1} obtained previously  \cite{Karch:2015rpa}. Here $\beta_0=\frac{\delta}{16\pi G_d}$, with $\delta$ an arbitrary dimensionless constant.

Equation (\ref{extendedholo1}) implies
\begin{equation}
\label{holosmarr1}
\alpha_k\frac{\partial X}{\partial\alpha_k}=\frac{d-2}{2(k-1)}g_k \frac{\partial X}{\partial g_k}
\end{equation}
for any arbitrary function $X$ of the parameters $\alpha_k$.  Setting $X=\Omega$, equation (\ref{holosmarr1}) then becomes
\begin{equation}
\alpha_k\frac{\partial\Omega}{\partial\alpha_k}=\frac{d-2}{2(k-1)}g_k \frac{\partial\Omega}{\partial g_k}
\end{equation}
After multiplying both sides by $2(k-1)$ and summing over $k$ we have 
\begin{equation}
\label{homo1}
\sum_{k=0}2(k-1)\alpha_k\Psi^{(k)}=(d-2)\sum_{k=0}g_k\frac{\partial\Omega}{\partial g_k} 
\end{equation}
where $\Psi^{(k)} = \frac{\partial\Omega}{\partial{\alpha_k}}$. We thus have the general relation
\begin{equation}
\label{homo3}
l\frac{\partial}{\partial l}+\sum_{k=1}2(k-1)\alpha_k\frac{\partial}{\partial{\alpha_k}}=(d-2)\sum_{k=0}g_k\frac{\partial}{\partial g_k}
\end{equation}
noting that  $-2\alpha_0\partial_{\alpha_0} =l\partial_l$.
Recalling that the Euler scaling relation $f(tx_1,\ldots,tx_m)=t^n f(x_1,\ldots, x_m)$ implies
\begin{equation}
n f(x_1,\ldots, x_m) = \sum_j x_j \frac{\partial f}{\partial x_j } 
\end{equation}
 for a  homogeneous function of order $n$, it is straightforward to see that equation (\ref{homo1}) can be written as 
\begin{equation}
\label{homo2}
\sum_{k=0}2(k-1)\alpha_k\Psi^k=(d-2)\Omega
\end{equation}
using
\begin{equation}
\Omega=\sum_{k=0}g_k\frac{\partial\Omega}{\partial g_k}
\end{equation} 
which   holds since $\Omega$ is   an homogeneous function of the $g_k$ of degree 1.

More generally $\Omega$ is a   function of $(g_k, R,Q)$ and not just the $g_k$.  
For any function $f(l, Z)$, its derivative with respect to $l$ will be
\begin{equation}
\partial_l f(l, Z)|_{Z_b}=\partial_lf|_Z+ p \frac{Z}{l}\partial_Z f|_l.
\end{equation}
if the quantity $Z$ has scaling behaviour $Z=Z_0 l^p$ for some constant $Z_0$.  For charged black holes, 
\begin{equation}
A_b=lA, ~~~~\mu_b=l\mu, ~~~~~Q_b=Q/l
\end{equation}
after converting to a canonical normalized field strength of dimension $2$, and the radius $R= R_0 l$
for the boundary CFT since 
\begin{equation}
ds^{2}_{boundary}=-dt^2+l^2d\Omega^{2}_{d-2}
\end{equation}
is the boundary metric \cite{Karch:2015rpa}.  Hence we obtain 

\begin{equation}
\label{homo5}
l\frac{\partial}{\partial l}+\sum_{k=1}2(k-1)\alpha_k\frac{\partial}{\partial{\alpha_k}}=(d-2)\sum_{k=0}g_k\frac{\partial}{\partial g_k}+R\frac{\partial}{\partial R}+Q\frac{\partial}{\partial Q}
\end{equation}
and so (\ref{homo1}) now reads as
\begin{eqnarray}
&&\sum_{k=0}2(k-1)\alpha_k\Psi^k\nonumber\\
&&=(d-2)\sum_{k=0}g_k\partial_{g_k}\Omega\big|_{\mu, T}+R\partial_R\Omega\big|_{\mu, T,\alpha_{k\geq 1}} + Q\partial_Q\Omega\big|_{\mu, T,\alpha_k}\nonumber\\
&&=(d-2)\Omega - M-\mu Q\nonumber\\
&&=(d-3)M-(d-2)TS-(d-3)\mu Q 
\label{smarr2}
\end{eqnarray}
upon using \eqref{Omega-eq}, which implies
\begin{equation}
d\Omega 
= -SdT-Qd\mu+vdp+\sum_{k\geq 1}\Psi^kd\alpha_k
\end{equation}
so that $\partial_Q\Omega=-\mu$ and $\partial_R\Omega = v \partial_Rp = -M $ from \eqref{equationofstate1} and \eqref{dRp}.
We see that \eqref{smarr2} is the Smarr relation \eqref{smarr}.  

\section{Some cases}

The main purpose of this section is to check the validity of the holographic equation of state (\ref{newstate}) for a variety of special cases. 

We shall consider spherically symmetric AdS Lovelolock black holes whose  metrics are given in (\ref{sphericalsym}) and \eqref{f-def}. The thermodynamic quantities for these black holes are given by  
\begin{eqnarray}
\tilde{M}&=&\frac{d-2}{16\pi G_d}\frac{1}{R^{d-2}}\sum^{\frac{d-1}{2}}_{k=0}\alpha_k\kappa^k r_{+}^{d-1-2k}+\frac{Q^2}{2(d-3)r_{+}^{d-3}R^{d-2}}\nonumber\\
T\tilde{S}&=&\frac{d-2}{4G_d}\frac{T}{R^{d-2}}\sum^{\frac{d-1}{2}}_{k=1}\frac{k\kappa^{k-1}\alpha_kr_{+}^{d-2k}}{d-2k}\nonumber\\
\mu\tilde{Q}&=&\frac{Q^2}{(d-3)r_{+}^{d-3}R^{d-2}}
\label{thermo-tilde}
\end{eqnarray}
and
\begin{eqnarray}
&&\tilde{\Psi}^{(k)}=\frac{d-2}{16\pi G_d}\kappa^{k-1}\frac{r_{+}^{d-2k}}{R^{d-2}}\big[\frac{\kappa}{r_+}-\frac{4\pi k T}{d-2k}\big]\nonumber\\
&& V=\omega^{(\kappa)}_{d-2}R^{d-2},~~~~~R=l.
\end{eqnarray}
From these equations the free energy density looks like
\begin{eqnarray}
\tilde{\Omega}&=&\frac{d-2}{16\pi G_d l^{d-2}}\sum_{k=0}\alpha_k\kappa^{k-1}r_+^{d-2k-1}\big[\kappa-\frac{4\pi k r_+ T}{d-2k}\big]\nonumber\\
&&\qquad - \frac{Q^2}{2(d-3)r_+^{d-3}l^{d-2}}
\end{eqnarray}
To compute $R\partial_R \tilde{\Omega}\big|_{N, \mu, T, \alpha_k^F}$ we have determine how the other quantities scale in term of $l$. It is easy to notice that $l^{d-2}/G_d\sim l^0, ~~r_+\sim l^2 T,~~ Q\sim l^{3d/2-4}T^{d-2},~~ \alpha_k=\alpha_k^Fl^{2(k-1)},~~~Q=lQ_b$~~ and so
\begin{eqnarray}
&&R\partial_R \tilde{\Omega}\big|_{N, \mu, T, \alpha_k^F} = l\partial_l\tilde{\Omega}\big|_{l^{d-2}/G_d, \mu, T,\alpha_k^F}\nonumber\\
&&= -\frac{d-2}{16\pi G_d l^{d-2}}\sum_{k=0}k\kappa^{k-1}\alpha_k r_+^{d-2k-1}\left(2\kappa - \frac{8\pi (k-1) r_+ T}{d-2k} \right)
\nonumber\\
\end{eqnarray}
and using \eqref{thermoquantities} and \eqref{thermo-tilde} it is also easy to check that 
\begin{eqnarray}
&&(d-1)\tilde{M}+R\partial_R\tilde{\Omega}\nonumber\\
&=&(d-2)\bigg[\frac{(d-2)T}{4G_dl^{d-2}}\sum_{k=0}\frac{k\kappa^{k-1}\alpha_k r_+^{d-2k}}{d-2k}+\frac{Q^2}{(d-3)r_+^{d-3}l^{d-2}}\bigg]\nonumber\\
&=& (d-2)(T\tilde{S}+\mu\tilde{Q})
\end{eqnarray}
recovering the equation of state  (\ref{equationofstate2}) with $n=d-2$.
 
We next consider rotating planar Lovelock black holes  in Gauss-Bonnet-Born-Infeld Gravity. The action  in $d$ dimensions is given by \cite{thermoofrot}
\begin{eqnarray}
\label{gaussbonet}
I_G&=&-\frac{1}{16\pi G_{d}}\int_{{\cal{M}}} d^{d}x\sqrt{-g}\big[ R-2\Lambda +\alpha (R_{\mu\nu\gamma\delta}\nonumber\\
&&\qquad R^{\mu\nu\gamma\delta}-4R_{\mu\nu}R^{\mu\nu}+R^2)
+ L(F)\big]\nonumber\\
&-&\frac{1}{8\pi G_{d}}\int_{\partial{\cal{M}}}\sqrt{-\gamma}\big[\Theta+ 2\alpha (J-2\hat{G}_{ab}\Theta^{ab})\big]
\end{eqnarray}
where $\Lambda=-(d-2)(d-1)/2l^2$ is the cosmological constant, $\alpha$ the Gauss-Bonnet coefficient and $L(F)$ the Born-Infeld Lagrangian
\begin{equation}
L(F)=4\beta^2\left(1-\sqrt{1+\frac{F^2}{2\beta^2}}\right)
\end{equation}
where $\beta$ is the Born-Infeld parameter which has a dimension of mass, $F^2=F^{\mu\nu}F_{\mu\nu}$ ~with~ $F_{\mu\nu}=\partial_\mu A_\nu-\partial_\nu A_\mu$.
Regarding the boundary term, $\Theta$ is the trace of extrinsic curvature $\Theta^{ab}$ of the boundary, $\hat{G}^{ab}(\gamma)$ is the Einstein tensor on the boundary and $J$ the trace of 
\begin{equation}
J_{ab}=\frac{1}{3}(\Theta_{cd}\Theta^{cd}\Theta_{ab}+2 \Theta \Theta_{ac}\Theta_b^c-2\Theta_{ac}\Theta^{cd}\Theta_{db}-\Theta^2\Theta_{ab})
\end{equation}
In order to solve the equations of motion derived from the action (\ref{gaussbonet}) we consider a  $d$ dimensional asymptotically AdS spacetime with $k$ rotation parameters, whose   metric reads \cite{lemos,awad}
\begin{widetext}
\begin{eqnarray}
\label{planar}
ds^2&=&-f(r)\bigg(\Xi dt-\sum^{k}_{i=1}a_i d\phi_i\bigg)^2+\frac{r^2}{l^2}\sum^{k}_{i=1}(a_i dt-\Xi l^2 d\phi_i)^2\nonumber\\
&+&\frac{dr^2}{f(r)}-\frac{r^2}{l^2}\sum^{k}_{i<j}(a_i d\phi_j-a_j d\phi_i)^2+ r^2 dX^2
\end{eqnarray} 
where $\Xi=\sqrt{1+\sum^{k}_{i}a_i^2/l^2}$ and $d X^2$ a $(d-2-k)$ dimensional Euclidean metric.
Using the ansatz
\begin{equation}
A_\mu=h(r)(\Xi \delta_\mu^0-\delta_\mu^ia_i)
\end{equation}
the equations of motion for the vector potential yield
\begin{equation}
h(r)=-\sqrt{\frac{d-2}{2(d-3)}}\frac{q}{r^{d-3}} {}_2F_1 \bigg(\frac{1}{2}, \frac{d-3}{2(d-2)}; \frac{3d-7}{2(d-2)}; -\eta\bigg)
\end{equation}
where ${}_2F_1(a, b; c;z)$ is a hypergeometric function and 
\begin{equation}
\eta=\frac{(d-3)(d-2)q^2}{2\beta^2 r^{2(d-2)}}
\end{equation}
Inserting (\ref{planar}) into the gravitational field equations  gives
\begin{equation}
f(r)= \frac{r^2}{2(d-4)(d-3)\alpha}(1-\sqrt{g(r)})
\end{equation}
where
\begin{eqnarray}
g(r)&=&1-16\frac{(d-4)\alpha\beta^2\eta}{d-1}{}_2F_1\bigg(\frac{1}{2}, \frac{d-3}{2(d-2)}; \frac{3d-7}{2(d-2)}; -\eta\bigg) \\
&&\qquad + 4\frac{(d-4)(d-3)\alpha}{(d-2)(d-1)r^{d-1}}\big(2\Lambda r^{d-1}+(d-2)(d-1)m-4\beta^2r^{d-1}(1-\sqrt{1+\eta})\big) 
\nonumber 
\end{eqnarray}
Setting $f(r_+)=0$,  from the above expression it follows that $m$ is given by
\begin{eqnarray}
m &=&-\frac{2\Lambda r_+^{d-1}}{(d-2)(d-1)}+\frac{4\beta^2 r_+^{d-1}}{(d-2)(d-1)}(1-\sqrt{1+\eta_+}) 
+ 2(d-2)^2\frac{q^2}{r_+^{d-3}}~{}_2F_1 \bigg(\frac{1}{2}, \frac{d-3}{2(d-2)}; \frac{3d-7}{2(d-2)}; -\eta_+\bigg) 
\nonumber
\end{eqnarray}
with $r_+$   the horizon radius.

  The thermodynamic quantities associated with these black holes are 
 
\begin{eqnarray}
M&=&\frac{1}{16\pi G_{d}} m ((d-1)\Xi^2-1)\nonumber\\
T&=&\frac{r_+}{2(d-2)\pi \Xi}[2\beta^2(1-\sqrt{1+\eta_+})-\Lambda]\nonumber\\
S&=&\frac{\Xi}{4G_{d}}r_+^{d-2}\nonumber\\
J_i&=&\frac{1}{16\pi G_{d}}(d-1) \Xi m a_i\nonumber\\
\omega_i&=&\frac{a_i}{\Xi l^2}\nonumber\\
Q&=&\frac{\sqrt{2(d-3)(d-2)}\Xi}{8\pi G_{d}}q\nonumber\\
\mu &=&\sqrt{\frac{d-2}{2(d-3)}}\frac{q}{\Xi r_+^{d-3}}{}_2F_1 \bigg(\frac{1}{2}, \frac{d-3}{2(d-2)}; \frac{3d-7}{2(d-2)}; -\eta_+\bigg)\nonumber\\
&&
\end{eqnarray}

\end{widetext}

For the $d$ dimensional black holes whose thermodynamic quantities given above we can see that 

\begin{eqnarray}
\label{planar1}
&&(d-1)\tilde{M}\nonumber\\
&&=\frac{1}{16\pi G_{d}l^{d-2}}(d-2)(d-1)m + \frac{ {(d-1)^2}}{16\pi G_{d}l^{d-2}} m \sum_i \frac{a_i^2}{l^2}\nonumber\\
&&=(d-2)(T\tilde{S}+\mu\tilde{Q})+ (d-1)\sum_i\omega_i\tilde{J}_i
\end{eqnarray}
which is (\ref{newstate}) upon setting $n=d-2$, provided $R\partial_R\tilde{\Omega}$ vanishes.

  To compute $R\partial_R\tilde{\Omega}$, we have to keep in mind that the bulk quantities $Q_b, \mu_b, J_i^b, \omega_i^b$ are redefined in the CFT as $Q=Q_bl,~\mu=\mu_b/l,~ J_i=J_i^b/l, ~\omega_i=l\omega_i^b$; we also have $l^{d-2}/G_{d}\sim l^0, ~~r_+\sim l^2 T, ~~ q\sim \mu l r_+^{d-3}$ and $a_i\sim l \omega_i$ and $R=l$. 
Hence a direct computation of \eqref{Omega-eq} yields
\begin{equation}
 \tilde{\Omega}=0
\end{equation}
and so
\begin{equation}
R\partial_R\tilde{\Omega}\big|_{N,\mu, T, \omega_i} = l\partial_l\tilde{\Omega}\big|_{l^{n-1}/G_{n+1},\mu, T, \omega_i} 
 = 0
\end{equation}
as expected.

 Finally we consider non-extremal rotating  black holes in minimal $5d$ gauged supergravity. This provides an interesting non-trivial example  with both charge and angular momentum. 
 The metric in the Boyer- Lindquist  coordinates $x^\mu=(t, r, \phi, \psi)$ reads \cite{generalnon} 
\begin{widetext}
\begin{eqnarray}
ds^2&=&-\frac{\Delta_\theta}{\Xi_a\Xi_b \rho^2}[(1+g^2 r^2)\rho^2 dt+ 2 q\nu]dt+\frac{2q}{\rho^2}\nu\zeta 
+ \frac{f}{\rho^4}\bigg(\frac{\Delta_\theta}{\Xi_a\Xi_b}dt-\zeta\bigg)^2+\frac{\rho^2}{\Delta_r}dr^2+\frac{\rho^2}{\Delta_\theta}d\theta^2\nonumber\\
&&\qquad + \frac{r^2+a^2}{\Xi_a}\sin^2\theta d\phi^2+\frac{r^2+b^2}{\Xi_b}\cos^2\theta d\psi^2\nonumber\\
A &=&\frac{\sqrt{3}q}{\rho^2}\bigg(\frac{\Delta_\theta}{\Xi_a \Xi_b}dt-\zeta\bigg)
\end{eqnarray}
where
\begin{eqnarray}
\nu &=& b\sin^2\theta d\phi + a\cos^2\theta d\psi \qquad\qquad ~~~
\zeta  =  a\sin^2\theta\frac{d\phi}{\Xi_a}+b\cos^2\theta\frac{d\psi}{\Xi_b} \nonumber\\
\Delta_\theta &=&1-a^2g^2\cos^2\theta -b^2g^2\sin^2\theta \qquad
\Delta_r  = \frac{(r^2+a^2)(r^2+b^2)(1+g^2r^2)+q^2+ 2abq}{r}-2m  \nonumber\\
\rho^2 &=& r^2+a^2\cos^2\theta+b^2\sin^2\theta \qquad
\Xi_a  = 1-a^2g^2 ~~~\Xi_b =1-b^2g^2   ~~~
f  =  2m \rho^2 - q^2+ 2abqg^2\rho^2
\end{eqnarray}
with $a, b$ ~the rotation parameters associated to the coordinates $\phi, \psi$ respectively and $g$ is a constant with dimension of length.
 
 The  associated thermodynamic quantities are 
\begin{eqnarray}
\label{minimal5d}
M&=&\frac{m\pi (2\Xi_a+2\Xi_b-\Xi_a\Xi_b)+ 2\pi q ab g^2 (\Xi_a+\Xi_b)}{4 \Xi_a^2\Xi_b^2 G_5}\nonumber\\
\omega_a &=&\frac{a(r_+^2+b^2)(1+g^2r_+^2)+bq}{(r_+^2+a^2)(r_+^2+b^2)+ abq}\nonumber\\
\omega_b &=&\frac{b(r_+^2+a^2)(1+g^2r_+^2)+aq}{(r_+^2+a^2)(r_+^2+b^2)+ abq}\nonumber\\
T &=&\frac{r_+^4[1+g^2(2r_+^2+a^2+b^2)]-(q+ab)^2}{2\pi r_+ [(r_+^2+a^2)(r_+^2+b^2)+ abq]}\nonumber\\
S &=&\frac{\pi^2 [(r_+^2+a^2)(r_+^2+b^2)+ abq]}{2\Xi_a\Xi_b r_+ G_5}\nonumber\\
J_a &=&\frac{\pi [2am+ qb(1+a^2g^2)]}{4\Xi_a^2\Xi_b G_5}\nonumber\\
J_b &=&\frac{\pi [2bm+ qa(1+b^2g^2)]}{4\Xi_a\Xi_b^2 G_5}\nonumber\\
Q &=&\frac{\sqrt{3}\pi}{4\Xi_a\Xi_b G_5}\nonumber\\
\mu &=&\frac{\sqrt{3}q r_+^2}{(r_+^2+a^2)(r_+^2+b^2)+ abq}
\label{sugratherm}
\end{eqnarray}
where $r_+$ is the horizon radius and
\begin{eqnarray}
 m=\frac{(r_+^2+a^2)(r_+^2+b^2)(1+g^2r_+^2)+q^2+2abq}{2r_+^2}\nonumber
\end{eqnarray}
\end{widetext}

In $5d$ equation (\ref{newstate}) takes the form
\begin{equation}
4\tilde{M}= 3(T\tilde{S}+\mu\tilde{Q})+ 4\omega_a \tilde{J}_a+4\omega_b\tilde{J}_b -R\partial_R\tilde{\Omega}
\end{equation}
To check the validity of this equation, we need to underline how the CFT quantities are defined in term of the bulk ones as in the previous example. $Q=Q^bl,~\mu=\mu^b/l,~ J_i=J_i^b/l, ~\omega_i=l\omega_i^b$. Also $l^3/G_5\sim l^0, ~~r_+\sim l^2 T, ~~ q\sim \mu l r_+^2$ and $a_i\sim l$~~~$(a_1=a,~ a_2=b)$ and $R=l$.  We compute $\tilde{\Omega}$ in the appendix and find from \eqref{app-dOm} that
\begin{eqnarray}
R\partial_R\tilde{\Omega}\big|_{N, T, \mu, \omega_i}&=& l\partial_l\tilde{\Omega}\big|_{l^3/G_5, T, \mu, \omega_i}\nonumber\\
&=&-4\tilde{M}+3(T\tilde{S}+\mu \tilde{Q})+4\omega_a\tilde{J}_a+4\omega_b\tilde{J}_b\nonumber\\
&&
\end{eqnarray}
showing that (\ref{newstate}) is satisfied.

\section{On the ~$g_k$(N)~dependence on $N$}

 In this section we investigate how the functions $g_k(N)$ can be approximated. To do so we employ the ansatz   that the $g_k(N)$ appear as   functions   of powers of $N^2$ \cite{Karch:2015rpa}.  More explicitly, we assume 
 \begin{equation}
\label{correspondance}
g_k(N)\equiv {\cal{ O}}(N^{2(1-k)}) 
\end{equation}
which means that for higher order curvature theories $(k\geq 2)$ the functions $g_k(N)$ are highly suppressed in the large $N$ field limit. We shall here restrict our attention to the leading terms of each higher curvature contribution. When we look at a higher curvature theory of gravity, the additional contributions   are seen as correction terms to the Einstein-Hilbert action.
\begin{figure}
\begin{subfigure}[b]{0.22\textwidth}
\centering
\begin{tikzpicture}[scale=1]
\draw (0,0) circle (1);
\draw (0,0) circle (1.5);
\end{tikzpicture}
\caption{•}\label{•}
\end{subfigure}
\begin{subfigure}[b]{0.22\textwidth}
\centering
\begin{tikzpicture}
\draw (0,0) circle (1.5);
\draw (5:1) arc (5:85:1);
\draw (95:1) arc (95:175:1);
\draw (185:1) arc (185:265:1);
\draw (275:1) arc (275:355:1);
\draw (0.1,0.97)--(0.1,-0.97);
\draw (-0.1,0.97)--(-0.1,-0.97);
\draw (-0.97,-0.1)--(-0.1,-0.1);
\draw (0.1,-0.1)--(0.97,-0.1);
\draw (0.1,0.1)--(0.97,0.1);
\draw (-0.1,0.1)--(-0.97,0.1);
\end{tikzpicture}
\caption{•}\label{•}
\end{subfigure}
\caption{(a) This planar diagram contributes to the scattering amplitude with an amount proportional to $N^2(g_{YM})^0=N^2\lambda^0$} and corresponds to $L^{(0)}$ in the Lovelock theory. As we know that every closed loop comes with a factor $N$.\\ 
(b) The non planar diagram, which is of great interest here, contributes to the scattering amplitude with a term proportional to $N^2 (g_{YM})^4=N^0\lambda^2$ and is linked to $L^{(1)}$. The Yang-Mills coupling $g_{YM}$ appears at each vertex of the diagram. 
\label{figure1}
\end{figure}
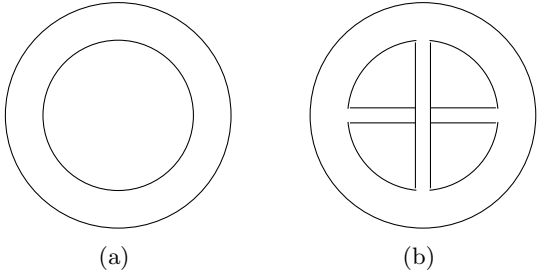
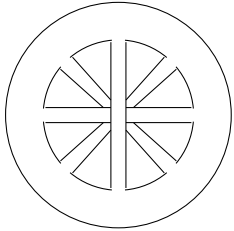
\begin{figure}
\begin{tikzpicture}
\draw (0,0) circle (1.5);
\draw (5:1) arc (5:40:1);
\draw (50:1) arc (50:85:1);
\draw (95:1) arc (95:130:1);
\draw (140:1)arc (140:175:1);
\draw (185:1) arc (185:220:1);
\draw (230:1)arc (230:265:1);
\draw (275:1) arc (275:310:1);
\draw (320:1) arc (320:355:1);
\draw (0.1,0.97)--(0.1,-0.97);
\draw (-0.1,0.97)--(-0.1,-0.97);
\draw (-0.97,-0.1)--(-0.1,-0.1);
\draw (0.1,-0.1)--(0.97,-0.1);
\draw (0.1,0.1)--(0.97,0.1);
\draw (-0.1,0.1)--(-0.97,0.1);
\draw (0.77,0.61)--(0.2,0.1);
\draw (-0.1,-0.2)--(-0.61,-0.77);
\draw (-0.2,0.1)--(-0.77,0.61);
\draw (0.1,-0.2)--(0.61,-0.77);
\draw (0.1,0.2)--(0.61,0.77);
\draw (-0.1,0.2)--(-0.61,0.77);
\draw (-0.2,-0.1)--(-0.77,-0.61);
\draw (0.2,-0.1)--(0.77,-0.61);
\end{tikzpicture}
\caption{The following diagram can be thought of as two similar copies of the four vertices non planar diagram piled together one on top of the other. It gives rise to a contribution to the scattering amplitude proportional to $N^2 (g_{YM})^8=N^{-2}\lambda^4$ and corresponds to $L^{(2)}$ in the Lovelock theory. For the term in $L^{(k)}$ in the Lovelock theory the corresponding diagram will consist in a stack of $k$ copies of the four vertices non planar diagram, whose the contribution to the scattering amplitude is proportional to $N^2 (g_{YM})^{2k}=N^{2(1-k)}\lambda^{2k}$ }
\label{figure2}
\end{figure} 

To better illustrate  what is meant, we can consider  a pure Yang-Mills theory, or  a field theory coupled to a Yang-Mills gauge theory,
 whose  Lagrangian $L^{(1)}$ gives rise to planar and non planar diagrams.    We can also consider a Kaluza-Klein-like model which couples gravity to a Yang-Mills gauge theory so that the general Lagrangian is similar to that of a pure gravity theory .   In such theories  gauge field self-interaction terms are part of the Lagrangian $L^{(1)}$, whereas  higher order diagrams come explicitly from  higher curvature terms ($R_{d+1}=R_d+gF^2+...$). 

Considering only  non-planar diagrams with 4 vertices, as shown in (\ref{figure1}), we know that these kinds of diagrams bring a contribution to the scattering amplitude proportional to $N^2g_{YM}^4=\lambda^2$, where $g_{YM}$ and $\lambda$ are the Yang-Mills and 't Hooft couplings respectively.  It clearly appears that these diagrams lead to contributions of the order of $N^0$ in the computation of the scattering amplitude. This reasoning generalizes to higher curvature terms $(k\geq 2)$ in the following way:
\begin{eqnarray}
&&L^{(0)}\sim {\cal{R}}^0~~\rightarrow ~~ N^2\nonumber\\
&&L^{(1)}\sim {\cal{R}}^1~~\rightarrow ~~ N^0\nonumber\\
&&L^{(2)}\sim {\cal{R}}^2~~\rightarrow ~~ N^{-2}\nonumber\\
&&L^{(k)}\sim {\cal{R}}^k~~\rightarrow ~~ N^{2(1-k)}
\end{eqnarray}
with ${\cal{R}}$ some scalar measure of the curvature.
Following the above construction we can infer that there should be a correspondence between the dependence of the functions $g_k(N)$ on $N^2$ and the contribution of non-planar diagrams to the scattering amplitude (\ref{figure2}), where we infer 
$N^2 (g_{YM})^{2k}=N^{2(1-k)}\lambda^{2k}$.  This suggests the correspondence
\eqref{correspondance}. These results are applicable to any higher-curvature theory of gravity.

\section{Conclusion}

By considering   the grand canonical free energy $\Omega$ in the both the bulk and the field theory on its boundary, we have  derived a holographic Smarr relation valid beyond the large $N$ limit, with subleading terms of the from $g_k(N)$ arising from higher-curvature corrections of the type found in $k$-th order Lovelock gravity.  By assuming   that $\Omega$ in the CFT is a homogeneous function of degree one of the $g_k(N)$ functions we were able to obtain the holographic equation of state \eqref{newstate}. We illustrated its validity  for several non-trivial cases in Lovelock gravity and in minimal gauged supergravity in 5 dimensions.
 
We expect  that asymptotically AdS black holes will in general satisfy the relations we have derived, testifying to the robustness of the correspondence between the bulk relations such as the Smarr relation and the  equation of state in the CFT. 
It was shown \cite{Karch:2015rpa} that  Einstein-gravity black holes  whose dual field theories are the large $N$ gauge theories with hyperscaling violation satisfy a modified equation of state in the large $N$ limit. We therefore expect that many other non trivial examples of black holes exist where the equation of state beyond this limit has a slightly modified form from the one we obtained. Some of these black holes are the black $Dn$ branes, which are dual to maximally supersymmetric gauge theories in $n+1$ dimensions.   Higher-curvature theories that are dual to gauge theories with hyperscaling violation should be good examples of such theories.  An interesting project for future work would be to investigate for these black holes the form of the equation of state at higher order.

\section*{Appendix}

We consider here a computation of the free energy density of   rotating black holes in minimal $5d$ gauged supergravity. From the thermodynamic quantities given in (\ref{minimal5d}) we  see that
\begin{eqnarray}
&& 4\tilde{M}-3(T\tilde{S}+\mu\tilde{Q})-4\omega_a \tilde{J}_a-4\omega_b\tilde{J}_b\nonumber\\
&=&\frac{\pi}{\Xi_a^2\Xi_b^2}m(2\Xi_a+2\Xi_b-\Xi_a\Xi_b)\frac{1}{G_5l^3}\nonumber\\
&+&\frac{2\pi}{\Xi_a^2\Xi_b^2}q ab g^2(\Xi_a+\Xi_b)\frac{1}{G_5l^3}-\frac{3}{G_5l^3}TS-\frac{3}{G_5 l^3}\mu Q\nonumber\\
&-&\frac{\pi}{l\Xi_a^2\Xi_b}[2am+qb(1+a^2g^2)]\frac{\omega_a}{G_5l^3}\nonumber\\
&-&\frac{\pi}{l\Xi_a\Xi_b^2}[2bm+qa(1+b^2g^2)]\frac{\omega_b}{G_5l^3}
\end{eqnarray} 
The free energy density reads as
\begin{widetext}
\begin{eqnarray}
\tilde{\Omega}&=&\frac{\pi m}{4\Xi_a^2\Xi_b^2G_5l^3}(2\Xi_a+2\Xi_b-\Xi_a\Xi_b)+\frac{\pi qabg^2(\Xi_a+\Xi_b)}{2\Xi_a^2\Xi_b^2G_5l^3} - \frac{\pi^2}{2\Xi_a\Xi_bG_5l^3r_+}T(r_+^2+a^2)(r_+^2+b^2)-\frac{\pi^2}{2\Xi_a\Xi_bG_5l^3r_+}Tabq\nonumber\\
&-&\frac{\sqrt{3}\pi}{4\Xi_a\Xi_bG_5l^3r_+}\mu ql-\frac{\pi am}{2\Xi_a^2\Xi_bG_5l^3r_+}\frac{\omega_a}{l}
-\frac{\pi bm}{2\Xi_a\Xi_b^2G_5l^3r_+}\frac{\omega_b}{l} - \frac{\pi qb(1+a^2g^2)}{4\Xi_a^2\Xi_bG_5l^3r_+}\frac{\omega_a}{l}-\frac{\pi qa(1+b^2g^2)}{4\Xi_a\Xi_b^2G_5l^3r_+}\frac{\omega_b}{l}
\end{eqnarray}
\end{widetext}

Therefore, we have
\begin{eqnarray}
&& l\partial_l\tilde{\Omega}\big|_{l^3/G_5, T, \mu, \omega_i}\nonumber\\
&&=\frac{\pi}{4\Xi_a^2\Xi_b^2}[-4m+ \frac{4\Xi_a\Xi_b}{\pi}TS+\frac{3}{r_+^2}(q^2+abq)]\nonumber\\
&&\qquad (2\Xi_a+2\Xi_b-\Xi_a\Xi_b)\frac{1}{G_5l^3}\nonumber\\
&&+(\frac{3}{2}-2)\frac{\pi}{\Xi_a^2\Xi_b^2}q ab g^2(\Xi_a+\Xi_b)\frac{1}{G_5l^3}+\frac{3}{G_5l^3}TS\nonumber\\
&&-\frac{\pi^2}{2\Xi_a\Xi_b r_+}[3r_+^4+(a^2+b^2)r_+^2-a^2b^2+2 abq]\frac{1}{G_5l^3}\nonumber\\
&&+\frac{\pi}{2l\Xi_a^2\Xi_b}[4am+2qb(1+a^2g^2)]\frac{\omega_a}{G_5l^3}\nonumber\\
&&+\frac{\pi}{2l\Xi_a\Xi_b^2}[4bm+2qa(1+b^2g^2)]\frac{\omega_b}{G_5l^3}\nonumber\\
&&-\frac{\pi}{2l\Xi_a^2\Xi_b}[\frac{4a\Xi_a\Xi_b}{\pi}TS+\frac{3a}{r_+^2}(q^2+abq)+\frac{3}{2}qb(1+a^2g^2)]\frac{\omega_a}{G_5l^3}\nonumber\\
&&-\frac{\pi}{2l\Xi_a\Xi_b^2}[\frac{4b\Xi_a\Xi_b}{\pi}TS+\frac{3b}{r_+^2}(q^2+abq)+\frac{3}{2}ab(1+b^2g^2)]\frac{\omega_b}{G_5l^3}\nonumber
\end{eqnarray}
and 
which can finally  be written as
\begin{eqnarray}\label{app-dOm}
&& l\partial_l\tilde{\Omega}\big|_{l^3/G_5, T, \mu, \omega_i}\nonumber\\
&=&-4\tilde{M}+3T\tilde{S}+4\omega_a\tilde{J}_a+4\omega_b\tilde{J}_b\nonumber\\
&+&\frac{9\pi}{4\Xi_a\Xi_b}\frac{q^2r_+^2}{[(r_+^2+a^2)(r_+^2+b^2)+abq]}\frac{1}{G_5l^3}\nonumber\\
&=&-4\tilde{M}+3(T\tilde{S}+\mu \tilde{Q})+4\omega_a\tilde{J}_a+4\omega_b\tilde{J}_b
\end{eqnarray}
using \eqref{sugratherm}.

\section*{Acknowledgments} This work was supported in part by
the Natural Sciences and Engineering Research Council of Canada.


\begin{thebibliography}{1}

\bibitem{thelarge} J. Maldacena, \emph{"The Large N Limit of Superconformal Field Theory and Supergravity,"} [arXiv:9711200v3~[hep-th]].
\bibitem{Kastor:2009wy} 
  D.~Kastor, S.~Ray and J.~Traschen,
  ``Enthalpy and the Mechanics of AdS Black Holes,''
  Class.\ Quant.\ Grav.\  {\bf 26}, 195011 (2009).  
\bibitem{Kubiznak:2012wp} 
  D.~Kubiznak and R.~B.~Mann,
  ``P-V criticality of charged AdS black holes,''
  JHEP {\bf 1207}, 033 (2012) 
\bibitem{Gunasekaran:2012dq} 
  S.~Gunasekaran, R.~B.~Mann and D.~Kubiznak,
  %``Extended phase space thermodynamics for charged and rotating black holes and Born-Infeld vacuum polarization,''
  JHEP {\bf 1211}, 110 (2012)
\bibitem{Altamirano:2013ane} 
  N.~Altamirano, D.~Kubiznak and R.~B.~Mann,
  ``Reentrant phase transitions in rotating anti?de Sitter black holes,''
  Phys.\ Rev.\ D {\bf 88}, no. 10, 101502 (2013)
  \bibitem{Altamirano:2013uqa} 
  N.~Altamirano, D.~Kubiz?ák, R.~B.~Mann and Z.~Sherkatghanad,
  ``Kerr-AdS analogue of triple point and solid/liquid/gas phase transition,''
  Class.\ Quant.\ Grav.\  {\bf 31}, 042001 (2014)
\bibitem{Hennigar:2016xwd} 
  R.~A.~Hennigar, R.~B.~Mann and E.~Tjoa,
  ``Superfluid Black Holes,''
  Phys.\ Rev.\ Lett.\  {\bf 118}, no. 2, 021301 (2017).
\bibitem{Kubiznak:2016qmn} 
  D.~Kubiznak, R.~B.~Mann and M.~Teo,
  ``Black hole chemistry: thermodynamics with Lambda,''
  Class.\ Quant.\ Grav.\  {\bf 34}, no. 6, 063001 (2017)
\bibitem{Johnson:2014yja} C. V. Johnson, \emph{"Holographic Heat Engines,"} Class. Quant. Grav. {\bf{31}} (2014) 205002 [arXiv:1404.5982].
  \bibitem{Dolan:2014cja} 
  B.~P.~Dolan,  ``Bose condensation and branes,''
  JHEP {\bf 1410} (2014) 179.
\bibitem{Kastor:2014dra} D. Kastor, S. Ray and Traschen, \emph{"Chemical Potential in the First Law for Holographic Entanglement Entropy,"} JHEP {\bf{11}}(2014)[1409.3521].

\bibitem{Karch:2015rpa} A. Karch and B. Robinson, \emph{"Holographic Black Hole Chemistry,"} [arXiv:1510.02472v2~[hep-th]].
\bibitem{Witten} E. Witten, \emph{"Anti-de Sitter space, thermal phase transition, and confinement in gauge theories,"} Adv. Theor. Math. Phys.{\bf 2}(1998)505-532 [arXiv:9803131~[hep-th]].
\bibitem{theeinstein} D. Lovelock, \emph{"The Einstein Tensor and its Generalizations,"} J. Math. Phys. {\bf{12}} (1971) 498-501.
\bibitem{smarrformulaand} D. Kastor, S. Ray and J. Traschen, \emph{"Smarr Formula and an Extended First Law for Lovelock Gravity,"} Class. Quant. Grav. {\bf{27}} (2010) 235014,  [arXiv:1005.5053].
\bibitem{multiple} A. M. Frassino, D. Kubiznak, R. B. Mann and F. Simovic, \emph{"Multiple Reentrant Phase Transitions and Triple Points in Lovelock Thermodynamics,"} [arXiv:1406.7015~[hep-th]].
\bibitem{isolatedcritical} B. P. Dolan, A. Kostouki, D. Kubiznak and R. B. Mann \emph{"Isolated Critical Point from Lovelock Gravity,"} [arXiv:1407.4783v1~[hep-th]].

\bibitem{phasetrans1} J. L. Zhang, R. G. Cai and H. Yu, \emph{"Phase Transition and Thermodynamical Geometry for Schwarzschild AdS Black Holes in $AdS_5~ S^5$ Spacetime,"} JHEP {\bf{02}} (2015) 143, [arXiv:1409.5305].
\bibitem{phasetrans2} J. L. Zhang, R. G. Cai and H. Yu, \emph{"Phase Transition and Thermodynamical Geometry of Reissner-Nordstrom-AdS Black Holes in Extended Phase Space,"} Phys. Rev. {\bf{D91}} (2015), no. 4 044028, [arXiv:1502.01428].
\bibitem{holoentang} E. Caceres, P. H. Nguyen  and J. F. Pedraza, \emph{"Holographic Entanglement Entropy and the Extended Phase Structure of STU Black Holes,"} [arXiv:1507.06069].
\bibitem{stringgenerated} D. G. Boulware and S. Deser, \emph{"String Generated Gravity Models,"} Phys. Rev. Lett.{\bf{55}} (1985) 2656.
\bibitem{symmetricsol1} J. T. Wheeler, \emph{"Symmetric Solutions to the Maximally Gauss-Bonnet Extended Einstein Equations,"} Nuclear Physics B {\bf{268}} (1986), no.3 737-746.
\bibitem{symmetricsol2} J. T. Wheeler, \emph{"Symmetric Solutions to the Maximally Gauss-Bonnet Extended Einstein Equations,"} Nuclear Physics B {\bf{273}} (1986), no.3 732-748.
\bibitem{anoteon} R. G. Cai, \emph{"A Note on Thermodynamics of Black Holes in Lovelock Gravity,"} Phys. Lett. {\bf{B582}} (2004) 237-242 [arXiv:0311240~[hep-th]].
\bibitem{blackholeentropy} T. Jacobson and R. C. Myers, \emph{"Black Hole Entropy and Higher Curvature Interactions,"} Phys. Rev. Lett. {\bf{70}} (1993) 3684-3687,  [arXiv:9305016~[hep-th]].
\bibitem{ontheuniversality} A. Castro, N. Dehmami, G. Giribert, and D. Kastor, \emph{"On the Universality of Inner Black Hole Mechanics and Higher Curvature Gravity,"} Journal of High Energy Physics {\bf{7}} (July, 2013) 164 [arXiv:1304.1696].
\bibitem{alovelockblack} X. O. Camanho and J. D. Edelstein, \emph{"A Lovelock Black Hole Bestiary,"} Class. Quant. Grav. {\bf{30}} (2013) 035009, [arXiv:1103.3669].
\bibitem{pathologies} T. Takahashi and J. Soda, \emph{"Pathologies in Lovelock AdS Black Branes and AdS/CFT,"} Class. Quant. Grav. {\bf{29}} (2012) 035008, [arXiv:1108.5041].
\bibitem{aspectsofholo} X. Dong, S. Harrison, S. Kachru, G. Torroba and H. Wang, \emph{"Aspects of holography for theories with hyperscaling violation,"} [arXiv:1201.1905v4~[hep-th]].
\bibitem{supergravityand} N. Itzhaki, J. M. Maldacena, T. Sonneschein and S. Yankielowicz \emph{"Supergravity and the Large N Limit of Theories with Sixteen Supercharges,"} [arXiv:9802042v3~[hep-th]].
\bibitem{thermoofrot} M. H. Dehghani and S. H . Hendi \emph{"Thermodynamics of Rotating Black Branes in Gauss-Bonnet-Born-Infeld Gravity,"} [arXiv:0611087v1~[hep-th]].
\bibitem{lemos} J. Lemos and V. Zanchin \emph{"Rotating Charged Black Strings and Three-Dimensional Black Holes,"} Phys. Rev. D {\bf 54}, 3840 (1996).
\bibitem{awad} A. M. Awad \emph{"Higher Dimensional Charged Rotating Solutions in (A)dS Space-times,"} Class. Quant. Grav. {\bf 20}, 2827 (2003)~[arXiv:0209238v3~[hep-th]].
\bibitem{generalnon} Z. W. Chong, M. Cvetic, H. Lu and C. N. Pope \emph{"General Non-Extremal Rotating Black Holes in Minimal Five-Dimensional Gauged Supergravity,"} ~~[arXiv:0506029v1~[hep-th]].
\end{thebibliography}
\end{document}